\documentclass[preprint,
 amsmath,amssymb,
 aps,
]{revtex4}

\usepackage{graphicx}
\usepackage{dcolumn}
\usepackage{bm}

\usepackage{tikz}
\usetikzlibrary{hobby}

\usepackage{float}
\usepackage{amsmath,amssymb}
\usepackage{pifont}
\usepackage{graphicx}
\usepackage{color}
\usepackage{array}
\newcolumntype{P}[1]{>{\centering\arraybackslash}p{#1}}
\usepackage{hyperref}

\usepackage{graphicx}
\usepackage{amsmath,amssymb,graphicx}
\usepackage{tabularx}
\newcolumntype{C}[1]{>{\centering\arraybackslash}p{#1}}
\usepackage{hyperref}
\usepackage{comment}
\usepackage{xcolor}
\usepackage{cleveref}
\newcommand{\be}{\begin{equation}}
\newcommand{\ee}{\end{equation}}
\newcommand{\bea}{\begin{eqnarray}}
\newcommand{\eea}{\end{eqnarray}}
\newcommand{\non}{\nonumber}

\begin{document}
\title{Vacuum decay and quadratic gravity: the massive case}
\author{Silvia Vicentini}

\email{silvia.vicentini@unitn.it}
\author{Massimiliano Rinaldi}%
\email{massimiliano.rinaldi@unitn.it}
\affiliation{Dipartimento di Fisica, Universit\`{a} di Trento,\\Via Sommarive 14, I-38123 Povo (TN), Italy}
\affiliation{Trento Institute for Fundamental Physics and Applications (TIFPA)-INFN,\\Via Sommarive 14, I-38123 Povo (TN), Italy}

\date{\today}

\begin{abstract}False vacuum decay in field theory may be formulated as a boundary value problem in Euclidean space. In a previous work, we studied its solution in single scalar field theories with quadratic gravity and used it to find obstructions to vacuum decay. For simplicity, we focused on massless scalar fields and false vacua with a flat geometry. In this paper, we generalize those findings to massive scalar fields with the same gravitational interactions, namely an Einstein-Hilbert term, a quadratic Ricci scalar, and a non-minimal coupling. We find that the scalar field reaches its asymptotic value faster than in the massless case, in principle allowing for a wider range of theories that may accommodate vacuum decay. Nonetheless, this hardly affects the viability of the bounce in the scenarios here considered. We also briefly consider other physically interesting theories by including higher-order kinetic terms and changing the number of spacetime dimensions.
\end{abstract}

\maketitle


\section{Introduction} 
\label{sec:intro}
Metastable states are separated from regions of lower energy by potential barriers with finite height and are classically stable at zero temperature. However, they can decay due to quantum tunneling. The decay rate of metastable fields depends on a solution of the Euclidean equations of motion called bounce  \cite{Coleman:1977py,Callan:1977pt,Coleman:1980aw}, which is usually assumed to have $O(4)$ symmetry. This is a trajectory between the tunneling point beyond the potential barrier and the false vacuum, which is reached at spacetime infinity. Once that the bounce is found, one can compute the decay rate from a combination of the on-shell action and a prefactor, called fluctuation determinant. There are several techniques, both numerical and analytical, which can be used to determine the bounce depending on the specific theory \cite{FerrazdeCamargo:1982sk,lee,Dutta:2011rc,Kanno:2012zf,Aravind:2014pva,Guada:2020ihz,Garfinkle:1989mv,Linde:1981zj,Claudson:1983et,Kusenko:1995jv,Kusenko:1996jn,Dasgupta:1996qu,Moreno:1998bq,John:1998ip,Masoumi:2016wot,Espinosa:2018hue,Espinosa:2018voj,Chigusa:2019wxb,Eckerle:2020opg,Isidori:2007vm,Salvio:2016mvj}. \\
 
 In a previous work \cite{nostro} we showed that the $O(4)$-symmetric bounce of a single scalar field theory is independent of the form of the potential as it approaches the false vacuum. This region is probed for large values of the radius of the  Euclidean spacetime. For simplicity, we required the false vacuum to have a flat geometry, and the scalar field to be massless, with small cubic self-interactions. Under these assumptions, the bounce can be determined analytically at large Euclidean radii. We named this solution the "asymptotic bounce".  This result was extended to include gravitational interactions, in the form of an Einstein-Hilbert term, a non-minimal coupling, and a quadratic Ricci scalar. By knowing the asymptotic bounce for the scalar field in all those cases we verified whether the following conditions on the bounce hold
 \begin{enumerate}
    \item the equations of motion have a solution such that all fields approach the false vacuum at  infinity;\item this solution has well-defined and finite on-shell action.
\end{enumerate} 
If any of the two conditions is violated we expect that the bounce solution does not contribute to vacuum decay. In this way, one can constrain vacuum decay in cosmological models with a scalar field and modified gravity.  The physical interpretation of Conditions 1. and 2. in terms of bubble formation was also considered. In particular, 
the violation of Condition 1. means that the only solution satisfying the bounce boundary conditions is the false vacuum static solution, and thus there is no phase transition. If, instead, the on-shell action is infinite and positive, ``vacuum quenching'' \cite{Coleman:1980aw} occurs. If infinite and negative, the semi-classical regime is broken. The bounce action may also be ill-defined at the upper bound of integration: this may happen, for example, if our candidate metastable state is a minimum of the Euclidean potential, and thus a maximum in Minkowski space \footnote{This result seems trivial, as there is no potential barrier through which the scalar field can tunnel. Actually, early studies of the vacuum decay phenomenon  focused on tunneling without barriers \cite{lee}, the most prominent example being  scalar field decay in a quartic potential with negative coupling}.  We found that quadratic gravity always forbids the bounce, while it is allowed in most cases when the field is non-minimally coupled to gravity. The aim of this paper is to consider a number of theories of great physical interest that, for simplicity, were disregarded from our previous analysis. One is, clearly, accounting for massive scalar fields. This may be particularly important for Higgs decay \cite{Degrassi:2012ry,Buttazzo:2013uya,Branchina:2014usa,Espinosa:2015kwx,Andreassen:2017rzq,Branchina:2013jra,Branchina:2019tyy,Bentivegna:2017qry,Andreassen:2016cvx,Branchina:2018xdh,Czerwinska:2016fky,Rajantie:2016hkj,Bentivegna:2017qry}. The Higgs mass is usually neglected in decay rate calculations but it might be important in the light of Conditions 1. and 2. The asymptotic bounce in the case of a single massive scalar field has already been found by \cite{Affleck:1980mp}, but, so far and to our knowledge, there are no generalizations that include gravitational contributions.  We also overlooked derivative self-interactions and scalar field decay in a number of space dimensions $d$ other than three. The former has been found to be a candidate high-energy correction to solve the hierarchy problem \cite{Shaposhnikov:2018xkv,Shaposhnikov:2018jag,Shkerin:2019mmu,Shaposhnikov:2020geh,Karananas:2020qkp}. The latter might be relevant as there are recent proposals for analogue experiments with $d<3$ \cite{Fialko:2014xba,Fialko:2016ggg,Braden:2017add,Braden:2018tky,Braden:2019vsw,Michel:2019nwa,Abed:2020lcf}. Also, it has been recently proposed that our four-dimensional Universe may live on a five-dimensional bubble \cite{Banerjee:2020wix,Banerjee:2019fzz,Banerjee:2018qey,Koga:2020jok}.\\

 To begin with, in Sec.\ref{sec:recap}, we delineate the notation and quickly summarize our previous results to provide both a first example of our method and some basic formulas that will be needed for the rest of the paper. In Sec.\ref{sec:scalarfield}-\ref{sec:spacedim} we compute the asymptotic bounce of the aforementioned theories. Then, we focus on massive scalar fields in four spacetime dimensions interacting with modified gravity. In particular, we will consider
\begin{itemize}
    \item  a single scalar field theory with Einstein-Hilbert gravity, a non-minimal coupling $\xi \phi^2 R$ and a quadratic term $\alpha R^2$ (Sec.\ref{sec:scaledep}). We consider separately the contribution of the last two terms in order to distinguish their effect in relation to the asymptotic behavior of the scalar field;
    \item  a single scalar field theory with non minimal coupling $\xi \phi^2 R$ and a quadratic term $\alpha R^2$ (Sec.\ref{sec:scaleinv}). 
\end{itemize}
We conclude in Sec.\ref{sec:concl} with some remarks.
 
\section{Vacuum decay of massless fields}
\label{sec:recap}
\begin{figure}
\centering
\mbox{
\begin{tikzpicture} [scale=2.8]
\draw (0,0.1) to [curve through={(0.3,0)..( 0.6,0.3)..(0.8,0.7)..(1.02,0.5)..(1.12,0)..(1.14,-0.1)..(1.3,-0.35)..(1.4,-0.38)..(1.6,-0.34)..(1.8,-0.17)}](2.1,0.3);
\draw[-stealth] (0,-0.8) to (0,1.3);
\draw[-stealth] (0,-0.01) to (2.5,0);

\node at (0.2,1.2) [scale=1] {$V(\phi)$};
\node at (2.45,-0.1) [scale=1] {$\phi$};
\node at (0.3,-0.12) [scale=1] {$\phi_{\rm fv}$};
\node at (1.45,-0.47) [scale=1] {$\phi_{\rm tv}$};
\node at (.85,0.82) [scale=1] {$\phi_{\rm top}$};
\end{tikzpicture}
}
\caption{Potential with a false vacuum state at $\phi_{\rm fv}$ and a true vacuum state at $\phi_{\rm fv}$. $\phi_{\rm top}$ marks the scalar field value at the top of the potential barrier.}
\label{fig:potential}
\end{figure}
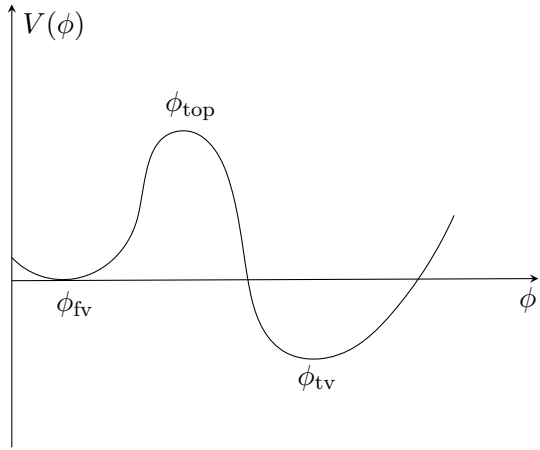

Here, we briefly summarize the results obtained in \cite{nostro} for massless scalar fields. Consider a scalar field theory in a four dimensional Euclidean spacetime described by the $O(4)$-symmetric action
 \begin{equation}\label{eq:action}S=2 \pi^2\int dt\, t^3 \left(\dfrac{\dot{\phi}^2}{2}+V(\phi)\right)\end{equation}
 with equation of motion 
 \begin{equation}\label{eq:eom1}\ddot{\phi}+\dfrac{3\,\dot{\phi}}{t}=\dfrac{dV}{d\phi}.\end{equation}
 Here $t$ indicates the four-dimensional Euclidean radius and the dot represent derivatives with respect to it.
  We choose $V(\phi)$ to have a false vacuum state $\phi_{\rm fv}$ such that $V(\phi_{\rm fv})=0$, which is separated by a potential barrier from regions where $V(\phi)<0$ (see Fig.\ref{fig:potential}). The false vacuum state decays through quantum tunneling beyond the potential barrier at an exponentially small rate $\Gamma$ given by
 \bea 
\Gamma=A\,e^{-B}.
\eea
 $B$ is the Euclidean action computed on 
a solution to the equation of motion, the bounce, that interpolates between a point beyond the potential barrier, which may be set at $t=0$ by translational invariance, and the false vacuum, at spacetime infinity. $A$ is the so-called quantum fluctuation determinant, which accounts for fluctuations around said trajectory. In the following, we label the bounce initial condition $\phi(0)$ as $\phi_0$, while $\phi_{in}$ indicates any other initial condition for the scalar field. Moreover, we set $\phi_{\rm fv}=0$, unless otherwise stated.\\
 
 In \cite{nostro} we determined the behavior of the bounce in the vicinity of the false vacuum. To do that, we considered a so-called undershoot trajectory. A scalar field is said to undershoot if it does not have enough energy to overcome the potential barrier in the region, as it evolves towards the false vacuum. It thus stops at some finite $t^*$, with $\phi_{\rm fv}<\phi(t^*)<\phi_{top}$ and starts oscillating around $\phi_{top}$. Such trajectory has initial condition $\phi_{in}<\phi_0$\cite{Coleman:1977py}. The closer is $\phi_{in}$ to $\phi_0$, the larger is $t^*$ and the closer is $\phi(t^*)$ to the false vacuum. In order to determine the asymptotic behavior of the bounce (or, for short asymptotic bounce) we expanded the equation of motion of the scalar field around $t^*$ for large $t$ as
 \bea
 \label{eq:taylorV}
 \ddot\phi+\dfrac{3\dot\phi}{t}=\sum_{n\geq 0}f_{n}\dfrac{(t-t^*)^n}{n!}\eea
 with 
 \bea
 f_n=\left(\dfrac{dV}{d\phi}\right)_*^{(n)}
 \eea
 and then let $t^*\rightarrow+\infty$. One finds that, if \bea
\label{eq:condV}
     \left(\dfrac{d^jV}{d\phi^j}\right)_* \ddot{\phi}_*^{j-2} t^{*2j-2}\ll1,\qquad \text{for $j\geq2$ and large $t^*$}
\eea 
the equation of motion may be approximated as
\bea 
\label{eq:eomapprox}\ddot{\phi}+\dfrac{3\dot{\phi}}{t}=\left(\dfrac{dV}{d\phi}\right)_*
\eea
i.e., the zeroth-order of the potential contribution dominates over the higher ones. Under these assumptions, Eq.\eqref{eq:eomapprox} gives 
 \bea\label{eq:behaviourtp}
\lim_{\underset{t^*\rightarrow+\infty}{\phi_*\rightarrow0}}\dot{\phi}(t)=-\dfrac{C_0}{t^3},\,\,\quad \lim_{\underset{t^*\rightarrow+\infty}{\phi_*\rightarrow0}}\left(\dfrac{dV}{d\phi}\right)_*=\dfrac{4 C_0}{t^{*4}},\,\,\quad  \lim_{\underset{t^*\rightarrow+\infty}{\phi_*\rightarrow0}}\phi(t)=\dfrac{C_0}{2 t^2},
\eea
 from  sufficiently large $t$ up to $t\rightarrow+\infty$. Using Eq.\eqref{eq:behaviourtp} we can recast condition Eq.\eqref{eq:condV}  as 
  \begin{equation}
 \label{eq:massless}
    \left( \dfrac{d^2V}{d\phi^2}\right)_* t^{*2}\ll1\,,\qquad 4 C_0\left(\dfrac{d^3V}{d\phi^3}\right)_*\ll1\,,
    \end{equation}
namely, the scalar field should be massless with small cubic self-interactions. \\

 This procedure may be easily extended to include Einstein-Hilbert gravity as long as the false vacuum has a flat geometry. The action is
 \bea\label{eq:actionEH}S=\int d^4x \sqrt{g} \left[-\dfrac{M_P^2}{2}R+\dfrac{1}{2} g_{\mu\nu} \nabla^{\mu}\phi\nabla^{\nu}\phi+V(\phi)\right] \eea
  We choose the line element to be $O(4)$ symmetric and given by
 \bea
 \label{eq:lineEH}
 ds^2=dt^2+\rho(t)^2 d\Omega_3^2.
 \eea
 The equations of motion are
 \begin{gather}\label{eq:eom1EH}\ddot{\phi}+3\,\dfrac{\dot{\rho}\,\dot{\phi}}{\rho}=\dfrac{dV}{d\phi}\\\label{eq:eom2EH}\dot{\rho}^2=1+ \dfrac{8 \pi}{3}\rho^2 \left(\dfrac{\dot{\phi}^2}{2}-V(\phi)\right).\end{gather}
 If the false vacuum has a flat geometry (that is $V(\phi_{\rm fv})=0$), one gets 
 \bea
 \label{eq:approxrho}
 \dfrac{\dot\rho}{\rho}\approx\dfrac{1}{t}
 \eea
 sufficiently near the bounce at large times and, thus, Eq.\eqref{eq:eom1EH} is the same as Eq.\eqref{eq:eom1}. Hence the calculation proceeds as before and  Eq.\eqref{eq:behaviourtp} holds.\\
 
By following a similar approach, we generalized our results to a modified gravity theory interacting with a scalar field. The action is
 \bea
 \label{eq:actionmod}
     S=2 \pi^2 \int_0^{+\infty}dt\,\rho(t)^3\left(\dfrac{\dot{\phi}^2}{2}+ V(\phi)-\dfrac{M_P^2}{2} R-\dfrac{\xi}{2} \phi^2 R+\dfrac{\alpha}{36} R^2\right).
 \eea 
 The equations of motion are 
 \begin{gather}
 \label{eq:eom1mod}\dot{\rho}^2=1+\rho^2 \, \dfrac{\dot{\phi}^2-2V(\phi)+\dfrac{\alpha}{36} R^2+\left(\dfrac{\alpha}{3}\dot{R}-12 \xi\,\phi\,\dot{\phi}\right)\dfrac{\dot{\rho}}{\rho}}{6\left(M_P^2+\xi \phi^2-\dfrac{\alpha}{18} R\right)},\\[5pt]
\label{eq:eom2mod}\ddot{\phi}+3\dfrac{\dot{\rho}\,\dot{\phi}}{\rho}=\dfrac{dV}{d\phi}-\xi \phi R.\end{gather}
The trace of the Einstein equations reads
 \bea
 \label{eq:tracemod}
    \left(3 M_P^2+ 3 \xi (1+6 \xi)\phi^2+\alpha\Box\right) R=3 \dot{\phi}^2 (1+6 \xi)+12 V(\phi)+18 \xi \phi \dfrac{dV}{d\phi}.
 \eea
 We separately analyzed the effect of non-vanishing $\xi$, $M_P^2$ and $\alpha$ on Conditions 1. and 2. by finding the asymptotic bounce with the method outlined above. More details on the calculation are reported in \cite{nostro}. Our results are summarized in Table \ref{tab:prevres}.
 \begin{table}[]
     \centering
     \begin{tabular}{|P{45mm}|c|c|}
     \hline
     Theory&Condition 1&Condition2\\
      \hline
        $M_p,\,\xi$  & \ding{51}&\ding{51} \\
         $M_p,\,\alpha$  & \ding{51}&\ding{55} \\
          $M_p,\,\xi,\,\alpha$  & \ding{51}&\ding{55} \\
       $\xi,\,\phi_{\rm fv}=0 $& \ding{55}&\ding{55} \\
       $\xi,\,\phi_{\rm fv}\neq0$, $\ell_{ml} \gg \phi_{\rm fv}$   & \ding{55}&\ding{55} \\ 
       $\xi,\,\phi_{\rm fv}\neq0$, $\ell_{ml} \ll \phi_{\rm fv}$   & \ding{51}&\ding{51} \\
        $M_p,\,\alpha$  & \ding{51}&\ding{55} \\
          $M_p,\,\xi,\,\alpha$  & \ding{51}&\ding{55} \\
       \hline
     \end{tabular}
     \caption{Summary of our previous results on the existence of the bounce through Conditions 1. and 2. for massless scalar fields interacting with quadratic gravity. In the left column, we indicate the theory considered by reporting the gravitational couplings $M_P,\,\xi,\,\alpha,$ which are taken to be non-vanishing and, possibly, some conditions on the false vacuum value $\phi_{\rm fv}.$ If Condition 1. does not hold there is no bounce-like trajectory and so we set also Condition 2. to be violated. $\ell_{ml}$ here indicates the width of the potential barrier.}
     \label{tab:prevres}
 \end{table}
\section{The asymptotic bounce of massive scalar fields}
 \label{sec:scalarfield}
The asymptotic bounce associated to a scalar field with mass $m$ and in the absence of gravitational interactions was found in \cite{Affleck:1980mp} and reads
\bea
\label{eq:bouncemass}
\phi(t)=C_0 e^{-m t}.
\eea
In such a case, the action Eq.\eqref{eq:action} diverges on the bounce if one uses Eq.\eqref{eq:behaviourtp}, as it gives
 \bea
 \int_0^{+\infty}dt \, t^3 \phi^2\approx \ln(t)\rvert_0^{+\infty}.\eea 
 The purpose of this Section is to find an alternative method to obtain Eq.\eqref{eq:bouncemass}, such that it can be easily generalized to include an Einstein-Hilbert term and also modified gravity models. To do that, we adopt a similar approach to the one outlined in Sect.\ref{sec:recap} by Taylor expanding the equation of motion in $t$ and determining which terms dominate at large radii on the bounce. Actually, Eq.\eqref{eq:bouncemass} suggests that
 \bea
 \dfrac{3\dot{\phi}}{t}\ll \ddot{\phi}\approx m^2 \phi\qquad \text{for} \qquad\phi(t)=C_0e^{-mt}
 \eea
 and, thus, the friction term (that is, the second term of Eq.\eqref{eq:eom1} should be included in the expansion and compared with the mass term. 
Then, we write the equation of motion as
 \bea
 \label{eq:eomnoder}
 \ddot{\phi}=m(t)^2 \phi-\dfrac{d}{dt}\left(\dfrac{3\phi}{t}\right)
 \qquad m(t)^2=m^2-\dfrac{3}{t^2}.\eea
 Defining
 \bea
 \dfrac{\partial V_1}{\partial \phi}\equiv m(t)^2 \phi,\qquad \dfrac{\partial V_2}{\partial \phi}\equiv -\dfrac{3\phi}{t},\eea
 we get
 \bea
  \ddot{\phi}=\dfrac{\partial V_1}{\partial\phi}+\dfrac{d}{dt}\left(\dfrac{\partial V_2}{\partial\phi}\right).
 \eea
To determine which terms dominate the Taylor expansion 
 \bea
\dfrac{\partial V_1}{\partial\phi}+\dfrac{d}{dt}\left(\dfrac{\partial V_2}{\partial\phi}\right)=\sum_{n\geq0} \left(\left(\dfrac{\partial V_1}{\partial\phi}\right)^{(n)}+\left(\dfrac{\partial V_2}{\partial\phi}\right)^{(n+1)}\right)\dfrac{(t-t^*)^n}{n!}
 \eea
on the bounce at large times we take sufficiently large $t$, $t^*$ by setting $t-t^*=-A t^*$ with $A$ some constant of order $O(0.1)$, and then let $t^*\rightarrow +\infty$.
Then one gets
 \bea
  \label{eq:taylormass}
  \sum_{n\geq0} \left(\left(\dfrac{\partial V_1}{\partial\phi}\right)^{(n)}+\left(\dfrac{\partial V_2}{\partial\phi}\right)^{(n+1)}\right)(t-t^*)^n\approx
 \sum_{n\geq0} \left(\dfrac{\partial V_1}{\partial\phi}\right)^{(n)}t^{*n}+ \sum_{n\geq1} \left(\dfrac{\partial V_2}{\partial\phi}\right)^{(n)} t^{*n-1}
 \eea
apart from numerical factors and shifting $n$ to $n-1$ in the second term. Then, Eq.\eqref{eq:taylormass} is tantamount to the Taylor expansion of a theory with equation of motion
 \bea
 \label{eq:approxeom}
 \ddot{\phi}=m(t)^2 \phi
 \qquad m^2(t)=m^2-\dfrac{3}{t^2}-\dfrac{3}{t\,t^*}
 \eea
 apart that the latter has an additional zeroth-order term. So, if we find the $t-$dependent part to be negligible with respect to the constant mass term at each order in the Taylor expansion for a theory as in Eq.\eqref{eq:approxeom}, we expect that the same holds for the original one. Defining
 \bea
 \dfrac{\partial V_3}{\partial \phi}=\dfrac{1}{t^*}\dfrac{\partial V_2}{\partial \phi},\eea
one gets
 \bea
 \left(\dfrac{\partial^{2+j}(V_1+V_3)}{\partial\phi^2\partial t^j}\right)_*t^{*j}\approx t^{*-2}\qquad\text{for  $j\geq 1$,}
 \eea
 and
 \bea
\left( \dfrac{\partial^{i+1} (V_1+V_3)}{\partial \phi\partial t^i}\right)_*t^{*i} \approx t^{*-2} \phi_*\qquad\text{for $i\geq 1$.}
 \eea
As reported in the Appendix, this means that the radius-dependent part of the potential is negligible with respect to the mass term one at each order in the Taylor expansion. Then, the equation of motion reduces to
\bea
\label{eq:eommass}
\ddot{\phi}\approx m^2 \phi
\eea
for large $t^*$ near the bounce. A solution to Eq.\eqref{eq:eommass} is
\bea
\label{eq:chap3:fullbouncemass}
\phi(t)=\dfrac{\phi_*}{2}e^{m t^*}( e^{-m t}+e^{-2 m t^*}e^{m t})
\eea
where integration constants are chosen such that $\dot{\phi}(t^*)=0$ and $\phi(t^*)=\phi_*.$ Thus, taking the limit $t^*\rightarrow+\infty$, Eq.\eqref{eq:chap3:fullbouncemass} reduces to
\bea
\label{eq:chap3:massddot}
\phi(t)=C_0 e^{-mt},\qquad \ddot{\phi}_*=m^2C_0 e^{-m t^*}.
\eea
For $\phi$ as in Eq.\eqref{eq:bouncemass}, energy is approximately conserved, as
\bea
\label{eq:bounceenergy}
\dfrac{\dot{\phi}^2}{2}-V(\phi)=0\qquad\text{for $\phi(t)=C_0 e^{-mt}$ and $V(\phi)=\dfrac{ m^2}{2} \phi^2$}.
\eea
One may think that, by adding a mass term to a  theory with a negative quartic potential, the field would overshoot for all $\phi(0)$, as the energy loss is reduced with respect to the $\lambda \phi^4$  solution. Actually, the field undershoots: the reason is that the scalar field decays as a power of $t$ even when the mass term dominates over the quartic one in the potential, but their contribution to the equation of motion is negligibly small with respect to the friction term and $\ddot{\phi}$. Such mass term induces additional loss of energy in the system, and, as a result, the field cannot climb the hill to reach the false vacuum. In this way, one recovers the well-known fact that a massive scalar field theory with a quartic potential does not have a bounce.\\

As reported in the previous section, including Einstein-Hilbert gravity just amounts to taking Eq.\eqref{eq:approxrho} near the bounce at large times and thus calculations are the same as the ones reported here.

\section{ Higher-order kinetic terms}
\label{sec:hokin}
Quantum tunnelling through an energy barrier has an exponentially small probability to occur in the semi-classical approximation. The smallness of some numbers, for example the ratio among the Higgs vacuum expectation value and the Planck mass, may be viewed in these terms (thus alleviating the hierarchy problem  \cite{Shaposhnikov:2018xkv,Shaposhnikov:2018jag,Shkerin:2019mmu,Shaposhnikov:2020geh,Karananas:2020qkp}), as 
\bea
\langle \phi\rangle=\int\mathcal{D}\phi\,\phi\,e^{-S}\equiv M_P e^{-W},
\eea
where $W$ is the generating functional computed on the bounce. The Coleman-de Luccia instanton is found by solving the equations of motion of the original theory with a pointlike source 
\bea
\phi(0)\equiv M_P \exp\left( M_P^{-1}\psi(0)\right)=M_P\exp\left(M_P^{-1}\int d^4 x \delta(x) \psi(x)\right).
\eea which generates a singular instanton at $t=0$. The singularity drives $W$ to infinity. It has been shown that a possible way to make it finite is to add higher order kinetic terms $
(\partial \psi)^n$, $n>2$, to the Lagrangian\cite{Shaposhnikov:2018xkv,Shaposhnikov:2018jag,Shkerin:2019mmu,Shaposhnikov:2020geh,Karananas:2020qkp}. Another application of higher-derivative terms to solve the hierarchy problem is addressed in \cite{Salvio:2020axm} regarding the agravity theory \cite{Salvio:2014soa}.\\

To find the asymptotic bounce, the right-hand side of the equation of motion in presence of higher-order kinetic terms 
\bea
\label{eq:eomhighor}
\ddot{\psi}+\dfrac{3\dot{\psi}}{t}=\dfrac{dV}{d\psi}+
n(n-1)\ddot{\psi}\dot\psi^{n-2}+\dfrac{3 n\dot{\psi}^{n-1}}{t}\eea
needs to be Taylor expanded around the turning point $t^*$. Here $V(\psi)$ contains non-derivative terms that generate a potential barrier, through which the scalar field can tunnel. 
Notice that $\ddot\psi\dot\psi^{n-2}$ may be written as 
\bea
\label{eq:nderi}
\left(\psi^{n-1}\right)^{(n)}=n!\dfrac{(n-1)}{2} \ddot\psi\dot\psi^{n-2}+\dots
\eea
This allows to consider the Taylor expansion of $\left(\psi^{n-1}\right)^{(n)}$ instead of $\ddot\psi\dot\psi^{n-2}$ and, if it satisfies Eq.\eqref{eq:condV}, also $\Box \psi\left(\partial \psi\right)^{n-2}$ does. In fact, our considerations for the Taylor expansion in the Appendix (and in the Appendix of \cite{nostro}) include all possible contributions to $f_n$ independently on numerical factors, and they are separately set to be small: this makes the calculation independent on incidental cancellations among $\ddot\psi\dot\psi^{n-2}$ and term in $\dots$ and thus, if higher-orders in the Taylor expansion of $\left(\psi^{n-1}\right)^{(n)}$ are found to be small, all terms on the right-hand side of Eq.\eqref{eq:nderi} are as such. Using
\bea\label{eq:approxexp}\sum_{j=1}\left(\psi^{n-1}\right)_*^{(n+i)}\dfrac{(t-t^*)^j}{j!}\underset{t-t^*\approx-At^*}{=}\sum_{j=n+1}\left(\dfrac{\psi^{n-1}}{t^{*n}}\right)_*^{(j)}\dfrac{t^{*j}}{j!},
\eea
as in Sect.\ref{sec:scalarfield}, one finds that Eq.\eqref{eq:approxexp} is tantamount to the Taylor expansion of a theory with equation of motion
\bea
\label{eq:approxeom2}
\ddot\psi+\dfrac{3\dot\psi}{t}\left(1-\dot\psi^{n-2}\right)=\dfrac{dV}{d\psi}+\dfrac{\psi^{n-1}}{t^{*n}}\qquad n>2
\eea
apart that the latter has additional $n$ orders.So, if we find that each order in the Taylor expansion for a theory as in Eq.\eqref{eq:approxeom2} is negligible with respect to the zeroth-one, then the same holds for our theory. The latter term satisfies Eq.\eqref{eq:condV} for $n>2$, $j\geq1$ and, thus, it does not affect significantly the Taylor expansion. Moreover, we can take
\bea
\dfrac{3\dot\psi}{t}\left(1-\dot\psi^{n-2}\right)\approx \dfrac{3\dot\psi}{t}
\eea
to lowest order near the bounce at large times. Then, massless $\psi$ satisfy Eq.\eqref{eq:behaviourtp} for large $t$ on the bounce if higher-order kinetic terms are added to the Lagrangian. If $\psi$ is massive, then Eq.\eqref{eq:bouncemass} holds instead. 

\section{Changing the number of spacetime dimensions}
 \label{sec:spacedim}
 The results of Sec.\ref{sec:recap}-\ref{sec:scalarfield} may be extended to a scalar field theory defined on a spacetime of arbitrary dimension $d+1$ ($d$ space dimensions). The $O(d+1)$-symmetric action is (here $\Gamma$ is the Euler Gamma Function)
  \bea
  \label{eq:actiond}
  S_E=\dfrac{2 \pi^{(d+1)/2}}{\Gamma\left(\frac{d+1}{2}\right)}\int dt\, t^d\left(\dfrac{\dot{\phi}^2}{2}+V(\phi)\right).
  \eea with equation of motion
  \bea
  \label{eq:eom1dim}
  \ddot{\phi}+\dfrac{d}{t}\dot{\phi}=\dfrac{dV}{d\phi}.
  \eea
The proof in the Appendix and in \cite{nostro} do not depend on $d$, as it only amounts to a numerical factor in Eq.\eqref{eq:eom1dim}. Eq.\eqref{eq:massless} instead does, as it is computed on the asymptotic bounce in three dimensions. If Eq.\eqref{eq:condV} holds, the equation of motion has a $d$ dependent solution that, in the limit of large $t^*$, gives
  \bea
  \label{eq:largetp}\lim_{\underset{t^*\rightarrow+\infty}{\phi_*\rightarrow0}}\dot{\phi}(t)=-\dfrac{C_0}{t^d},\qquad \lim_{\underset{t^*\rightarrow+\infty}{\phi_*\rightarrow0}}t^{*d+1}\left(\dfrac{dV}{d\phi}\right)_*=(d+1)C_0.\eea
Eq.\eqref{eq:condV} and Eq.\eqref{eq:largetp} suggest that, for $d>3$, the scalar field should be massless. In $d=2$ instead they give
\bea
\label{eq:cond2d}
\left(\dfrac{d^2V}{d\phi^2}\right)_*t^{*2}\ll1,\qquad 
4 C_0\left(\dfrac{d^3V}{d\phi^3}\right)_*t^*\ll1, \qquad 16 C_0^2 \left(\dfrac{d^4V}{d\phi^4}\right)_*\ll1,
\eea
so, in this case, also cubic terms are excluded.
In $d=1$ instead one has that $\phi(t)$ diverges as $t\rightarrow +\infty$ (as can be seen by integrating the first of Eq.s\eqref{eq:largetp}) and thus higher-order terms in the Taylor expanded potential are important in determining the asymptotic bounce.\\
 
 In all the cases analyzed above, massive scalar fields are excluded, as they violate Eq.\eqref{eq:condV}, dominating the Taylor expansion at each order. The proof in Sect.\ref{sec:scalarfield} is independent of the number of spacetime dimensions, as it only amounts to a numerical factor in Eq.\eqref{eq:eom1dim}. So, if the potential is dominated by the mass term near the false vacuum, the asymptotic bounce is given by Eq.\eqref{eq:bouncemass}.\\

Including Einstein-Hilbert gravity does not change our results. The line element is
\bea
ds^2=dt^2+\rho(t)^2 d\Omega_d^2
\eea
and the equation of motion for the scale factor is as in Eq.\eqref{eq:eom2EH} for $d\geq 2$, with the factor $3$ replaced with a $d-$dependent term. Thus, at large times on the bounce, we get
  \bea
 \dot{\rho}^2=1+O(t^{2-2d})\qquad\text{for $\rho(t)\approx t$ and $\phi(t)\propto t^{-d+1}$}
  \eea
  for massless scalar fields and
    \bea
 \dot{\rho}^2=1+O(t^2 e^{-2 m t})\qquad\text{for $\rho(t)\approx t$ and $\phi(t)\propto e^{-m t}$}.
  \eea
  which consistently gives $\dot{\rho}\approx 1$ for $t\rightarrow +\infty$ for $d\geq 2$. In both cases the on-shell action is convergent as regards the upper bound of integration.

\section{The asymptotic bounce in modified gravity}
 \label{sec:scaledep}
 In this Section, we focus on massive scalar fields interacting with modified gravity, and test for Conditions 1. and 2. We adopt the same approach of our previous work, so some calculations (the ones independent on the scalar field mass) are identical. For brevity those steps are reported only in a summarized form (referring to \cite{nostro} for details),  and the focus instead is on parts that rely on the new results given above. The action is as in Eq.\eqref{eq:actionmod} and the equations of motion are Eq.s \eqref{eq:eom1mod}-\eqref{eq:tracemod}.
The scalar field potential $V(\phi)$ is such that 
 \bea
 V(\phi)\approx \dfrac{m^2}{2}(\phi-\phi_{\rm fv})^2
 \eea
 for $\phi$ near $\phi_{\rm fv}$. Unless explicitly stated, we take $\phi_{\rm fv}=0$. In the following, we separately analyze the effect of couplings $\xi$, $\alpha$ on the asymptotic bounce and  consider analogies and differences with respect to the massless case. In Sec.\ref{sec:scaleinv} the Einstein-Hilbert term is turned off and analogous calculations are performed. 
 
\subsection{Non-minimal coupling}
In this Section, we consider non-minimal coupling corrections to a single scalar field theory with Einstein-Hilbert gravity, namely, we set $\xi\neq 0$, $M_P\neq 0$, $\alpha=0$. As described above, the mass term in the potential dominates over the friction one when $\xi=0$. In order to find the asymptotic bounce, we recast the equation of motion in a similar form to Eq.\eqref{eq:eom1}, as we did for massless fields in \cite{nostro}. 
Near the bounce at large times the scalar field is close to $\phi_{\rm fv}=0$. Using Eq.\eqref{eq:tracemod} in the right-hand side of Eq.\eqref{eq:eom2mod} and taking
\bea
V(\phi)\approx\dfrac{m^2}{2 }\phi^2
\eea
one finds at lowest order
 \bea
\label{eq:expW}
\ddot{\phi}+\dfrac{3\dot{\phi}}{t}=\dfrac{dW}{d\phi}\quad\text{with}\quad \dfrac{dW}{d\phi}\approx m^2 \phi+\xi \phi\dot{\phi}^2 (1+6 \xi).
\eea
As we saw in Section \ref{sec:scalarfield}-\ref{sec:hokin}, we can consider the Taylor expansion of $(\phi^3)^{(2)}$ in place of $\phi\dot{\phi}^2$. If the former gives a negligible contribution to the Taylor expansion near the bounce at large times, also the latter does. Using
\bea\sum_{i=1}\left(\phi^{3}\right)_*^{(2)}\dfrac{(t-t^*)^i}{i!}\underset{t-t^*\approx-At^*}{=}\sum_{i=2}\left(\dfrac{\phi^{3}}{t^{*2}}\right)_*^{(i)}\dfrac{t^{*i}}{i!}
\eea
 one finds that the this term behaves as a quartic interaction in the potential. It is thus negligible with respect to $m^2 \phi$ and then the scalar field should satisfy Eq.\eqref{eq:bouncemass} \footnote{While we considered only $\phi_{\rm fv}=0$ here, the generalization to $\phi_{\rm fv}\neq 0$ is straightforward and it amounts to a rescaling of the potential at large times on the bounce \cite{nostro}. One finds that the asymptotic bounce is $\phi(t)=C_0 e^{-C_1 t}$ with $C_1=\left(1-\dfrac{6 \xi^2 \phi_{\rm fv}^2}{M_P^2+\xi (1+6\xi \phi_{\rm fv}^2)}\right) m$.}.\\

Taking the asymptotic bounce and $\rho(t)\approx t$ in Eq.\eqref{eq:eom1mod} gives consistently $\dot{\rho}\approx 1$, as long as $M_P$ is finite, so Condition 1. is satisfied. Moreover, Eq.\eqref{eq:tracemod} gives
\begin{gather}
R\approx \dfrac{9 C_0^2 m^2 e^{-2 m t}}{M_P^2}
\end{gather}
to lowest order on the bounce at large times. Then, the Lagrangian decays sufficiently fast so that its integral Eq.\eqref{eq:actionmod} converges in the upper bound. Thus, we cannot exclude a bounce (and, then, vacuum decay) in massive scalar field theories with Einstein-Hilbert gravity and a non-minimal coupling, in analogy to what was found in the massless case.

 \subsection{Quadratic gravity}
 In this Section we set $\alpha\neq 0,\,M_P^2\neq 0,\,\xi=0$. Massless scalar fields forbid a bounce when quadratic corrections  in $R$ are included \cite{nostro}. In particular, it was found that the friction term in Eq.\eqref{eq:tracemod} may be approximated as
 \bea
 \dfrac{\dot{\rho}}{\rho}\approx \dfrac{1}{t}
 \eea
 on the bounce at \emph{all} times, which allows to explicitly solve for $R$ independently of the scalar field potential and find
  \bea
     R&=&\epsilon^2C_1\,\dfrac{J_1\left(\epsilon\, t'\right)}{t}-\epsilon^2\dfrac{ \pi}{2}\dfrac{J_1\left(\epsilon\,t'\right)}{t}\int^t F(y) Y_1\left(\epsilon^3\,y'\right) y^2 dy\\\non
     &+&\epsilon^2\dfrac{\pi}{2}\dfrac{Y_1\left(\epsilon^3\,t'\right)}{t}\int^t F(y) J_1\left(\epsilon\,y'\right) y^2 dy\,.
     \eea
     Here, $J,Y$ are Bessel function of the first kind,
 \bea
 A=\sqrt{\dfrac{3M_{\rm P}^2}{|\alpha|} }\,,\quad t'=A t\,,\quad \begin{cases} \epsilon=1&\alpha>0\\\epsilon=i&\alpha<0\,, \end{cases}
 \eea
$i$ is the imaginary unit, $C_{1,2}$ are constants, and the function $F(t)$ is given by
 \bea
 F(t)=\dfrac{3\dot{\phi}(t)^2}{\alpha}+\dfrac{12V(\phi(t))}{\alpha} \,.
 \eea
 On the bounce at large times the scalar field is independent on $R$ and given by Eq.\eqref{eq:bouncemass}. Using  the asymptotic forms of the Bessel functions $J$, $Y$ for large arguments \cite{abraham} the Ricci scalar can be approximated as
\bea
\label{eq:Rlargetnonm}
R=\epsilon^2(C_1+\tilde{C}_1)\,\dfrac{J_1\left(\epsilon\, At\right)}{t}+\tilde{C}_2\dfrac{Y_1\left(\epsilon^3\, At\right)}{t}+O(e^{-2 mt}).
\eea
By the same argument in \cite{nostro}, one has that $C_1+\tilde{C}_1\neq0, C_2\neq 0$ on the bounce. Then, the Ricci scalar indeed reaches $R=0$  as $\phi$ approaches $\phi_{\rm fv}$ for $\alpha >0$ (it does not for $\alpha<0$), but the on-shell action is ill-defined. Then, Condition 2. is violated and a bounce is excluded in theories with an Einstein-Hilbert term and a quadratic one.

 \subsection{Quadratic gravity and non-minimal coupling}
 \label{sec:nonmflat}
 We now set both $\alpha\neq 0$, $\xi\neq 0$ which makes Eq.\eqref{eq:eom2mod} and Eq.\eqref{eq:tracemod} entangled through the non-minimal coupling. To find the asymptotic bounce, we use Eq.\eqref{eq:eom2mod} to replace non-derivative terms in $R$ in Eq.\eqref{eq:tracemod}, to find the full solution when Eq.\eqref{eq:approxrho} holds. The Ricci scalar is
  \bea
 \label{eq:solnonm}
 R=C_1+\dfrac{C_2}{2t^2}+\int^t t'^{-3} \int^{t'} F(t'') t''^{3} dt''dt'
 \eea
 with
 \bea
\alpha  F(t)=3 \dot{\phi}^2 (1+6 \xi)+12 V(\phi)+18 \xi \phi \dfrac{dV}{d\phi}-3 \left( (1+6 \xi) \phi+\dfrac{M_P^2}{\phi}\right)\left(\ddot{\phi}+3\dfrac{\dot{\phi}}{t}-\dfrac{dV}{d\phi}\right).
      \eea
We define $f(t)$ as
\bea
\label{eq:defft}
\int^t t'^{-3} \int^{t'} F(t'') t''^{3} dt'dt''\equiv f(t) F(t).
\eea
and replacing $R$ in Eq.\eqref{eq:eom2mod} with Eq.\eqref{eq:solnonm} one gets
 \bea
 \label{eq:eommix}
 \ddot{\phi} F_1(\phi,t)+\dfrac{3\dot{\phi}}{t} F_2(t)=V'(\phi)F_3(t)+12 \xi \phi V(\phi) f(t)
 \eea
 with
 \bea
 F_1(\phi,t)=1+\dfrac{3\xi}{\alpha
} \phi f(t) \left( (1+6 \xi) \phi+\dfrac{M_{\rm P}^2}{\xi\phi}\right)\approx \dfrac{3}{\alpha
}  M_{\rm P}^2 f(t) +1
 \eea
  \begin{gather}
 F_2(\phi,t)=1-\dfrac{3\xi}{\alpha
}  (1+6\xi)t \phi \dot{\phi} f(t)+\dfrac{3\xi}{\alpha
} \phi f(t) \left( (1+6 \xi) \phi+\dfrac{M_{\rm P}^2}{\xi\phi}\right)\approx \dfrac{3}{\alpha
}  M_{\rm P}^2 f(t)+1\notag\\
 \end{gather}
   \bea
   F_3(\phi,t)=1+\dfrac{18\xi^2}{\alpha
} \phi^2 f(t)+\dfrac{3\xi}{\alpha
} \phi f(t)\left( (1+6 \xi) \phi+\dfrac{M_{\rm P}^2}{\xi\phi}\right)\approx  \dfrac{3}{\alpha
}  M_{\rm P}^2 f(t) +1
 \eea
 near the bounce at large times. Therefore, if $f(t)\neq -\dfrac{\alpha}{3 M_P^2}$ the scalar field equation of motion may be approximated as 
 \bea
 \label{eq:chap4:approxphieom}
 \ddot{\phi}+\dfrac{3 \dot{\phi}}{t}=\dfrac{dV}{d\phi}
 \eea
 for small $\phi$, as if the non-minimal coupling was negligible. Thus  Eq.\,\eqref{eq:bouncemass}  hold. Then, we can plug this solution in Eq.\,\eqref{eq:tracemod} and Taylor-expand the potential and the friction term around the turning point $t^*$, in order to determine which terms dominate the equation of motion at large times on the bounce. One finds that Eq.\,\eqref{eq:timecond2} (in which $\phi$ is to be replaced by $R$ and $\rho$ by $t$) holds for $R_*\gg  e^{-2 m t_*}$. This gives, near the bounce at large times,
 \bea
 \label{eq:approxtrace}
 \alpha \ddot{R}\approx -3 M_P^2 R.
 \eea
 Thus, according to Eq.\eqref{eq:chap3:massddot}, $R$ decreases exponentially in $t$
 \bea
 \label{eq:approxR}
 R\approx C_0 e^{-\sqrt{\frac{3}{\alpha}} M_P t}.
 \eea
 Anyway $e^{-2mt}\gg e^{-M_P t}$ for $m\ll M_P$ and so Eq.\eqref{eq:approxtrace} is not a consistent approximation of Eq.\,\eqref{eq:tracemod}. This means that $\phi$ terms in Eq.\,\eqref{eq:tracemod}, as well as the Einstein-Hilbert term, are important at each order in the Taylor expansion. Then $R$ is given by Eq.\,\eqref{eq:Rlargetnonm}, as $\phi^2 R$ in Eq.\,\eqref{eq:tracemod} is negligible with respect to $R$ when $\phi$ is given by the asymptotic bounce. Constants $C_1+\tilde{C}_1$ and $C_2$ determine the behaviour on and near the bounce. If they vanish, $R$ is completely set by the scalar field and changing the initial condition $R(0)$ away from the bounce one still gives a bounce. Thus, this should correspond to the $\alpha=0$ limit and, for $\alpha \neq 0$, one has $C_1+\tilde{C}_1\neq0$ and $C_2\neq0$. If this is the case though, Eq.\eqref{eq:chap4:approxphieom} is not a consistent approximation of Eq.\,\eqref{eq:eom2mod}. This excludes a bounce for massive scalar fields\footnote{If instead $f(t)\approx  -\dfrac{\alpha}{3\xi M_P^2}$ on the bounce at large times, $R$ is given by Eq.\,\eqref{eq:approxR} which implies again that Eq.\,\eqref{eq:approxtrace} holds.  This makes the approximation Eq.\eqref{eq:chap4:approxphieom} again reliable, as $\phi R$ gives a negligible contribution in the Taylor expansion. As a result, $\phi$ terms dominate the Einstein-Hilbert one for massive scalar field, which is in conflict with previous statements.}.

\section{Scale-invariant gravity}
\label{sec:scaleinv}
In the previous section, it was found that that a quadratic Ricci term alone forbids a bounce according to Conditions 1. and 2., and this is primarily due to the Einstein-Hilbert term in the Lagrangian. Moreover,  this result is independent on the scalar field asymptotic bounce, as it enters only at higher orders in the large $t$ limit. For this reason we now make the gravitational sector scale invariant, taking $M_P=0$, and repeat the calculations in all three cases above considered.  
\subsection{Non-minimal coupling}
\begin{figure}
        \centering
         \includegraphics[scale=0.7]{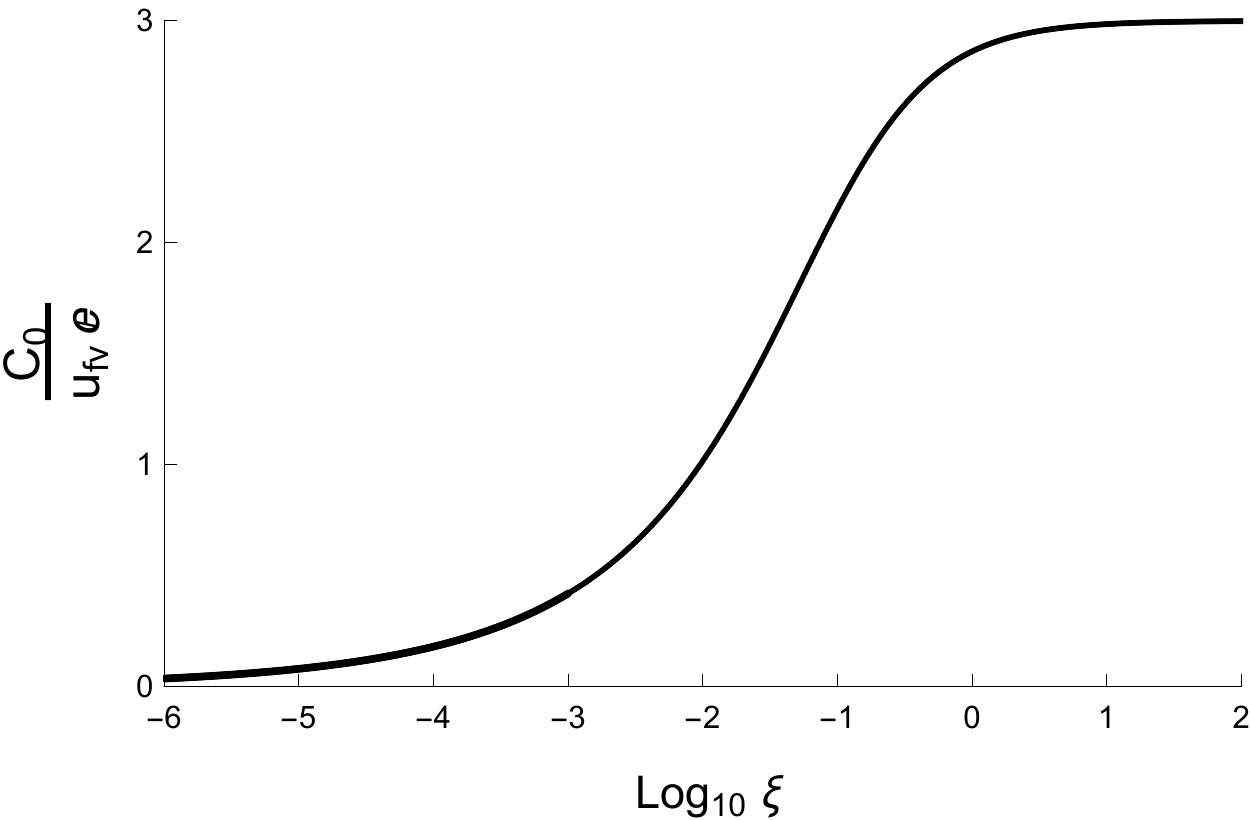}
        \caption{Zeros of $F(x,\xi)$ as a function of $\xi$.}
        \label{fig:gplot}
    \end{figure}

To find the asymptotic bounce, we use Eq.\eqref{eq:tracemod} in Eq.\eqref{eq:eom2mod} with $\alpha=M_P^2=0$. Replacing $\phi$ with $u=\phi^2$ the equation of motion can be casted as
  \bea\label{eq:fieldsquared}
  \ddot{u}+3\dfrac{\dot{\rho}\,\dot{u}}{\rho}=\dfrac{dW}{du}\equiv\dfrac{4}{1+6\xi}\left(u \dfrac{dV}{du}-2 V(u)\right).\eea Retaining the mass term only one gets
  \bea
  \dfrac{dW}{du}=-\dfrac{2}{6 \xi+1} m^2 u.
  \eea
  Thus, if $6 \xi+1<0$, 
  \bea
  \phi(t)=C_0 e^{-C_1 t},\qquad \text{where $C_1=\sqrt{-\dfrac{2 m^2}{6\xi+1}}$}.
  \eea
  Plugging this solution in Eq.\eqref{eq:eom1mod} one finds that there is an inconsistency in the boundary conditions for gravity, as using $\rho(t)=t$ and Eq.\eqref{eq:bouncemass} gives
  \bea\label{eq:wrongrhosi}
  \dot{\rho}^2=3+\left(\dfrac{m^2}{12\xi (1+6\xi)}-\dfrac{m^2}{6 \xi}\right) t^2+O\left(t\right).
  \eea Thus, there is no bounce if the false vacuum has a flat geometry. This situation improves if the scalar field has a non-vanishing vacuum value. In fact, taking $V(\phi_{\rm fv})=0$ and $\phi_{\rm fv}\neq 0$, we have
  \bea
  \label{eq:dWdufv}
  \left(u \dfrac{dV}{du}-2 V(u)\right)\approx \dfrac{m^2}{4} (u-u_{fv}) 
  \eea
  where $u_{fv}=\phi_{\rm fv}^2$. Thus
  \bea
  \label{eq:bouncefv}
  \phi(t)\approx\sqrt{u_{fv}+C_0 e^{-C_1 t}}\approx\sqrt{u_{fv}}+\dfrac{C_0}{2 \sqrt{u_{fv}}} e^{-C_1 t}\quad\,\text{  with}\quad C_1=\sqrt{\dfrac{m^2}{1+6 \xi}}
  \eea 
  for sufficiently large $t$. The transition to the asymptotic regime (say occuring at some $t=\bar{t}$) should be reached only in a narrow region around $u_{fv}$,
\bea
\dfrac{C_0e^{-C_1 \bar{t}}}{u_{fv}}\ll 1,
\eea
as otherwise the abovementioned inconsistencies in Eq.\eqref{eq:eom1mod} occur. The values of $C_0$ and $\bar{t}$ are theory-dependent and the combination
\bea
\ell_m\equiv C_0e^{-C_1 \bar{t}},
\eea
marks the distance of $\phi$ from $\phi_{\rm fv}$ at which the mass term dominates the potential. One gets
  \bea
  \dot{\rho}(t)=1+O\left(t^2 e^{-2 C_1 t}\right)
 \eea
for $\rho(t)= t$ and the scalar field given by the asymptotic bounce, thus satisfying Condition 1. If $\bar{t}$ is sufficiently small though, the scalar field is approximately given by Eq.\eqref{eq:bouncemass}, which possibly makes the second term in Eq.\,\eqref{eq:eom1mod} large and negative, making $\dot\rho$ vanish at finite times for some values of $C_0, \, u_{fv},\,\xi,\,m$. However, in order to avoid quantum gravity effects, all scales should be (much) smaller than the Planck mass, which gives, according to Eq.\,\eqref{eq:eom1mod}, $\dot{\rho}\approx 1$ on the bounce at \emph{all} times. To see this we take Eq.\eqref{eq:eom1mod} and impose
\bea
\dot{\rho}=0, \qquad \rho=t,\qquad\phi(t)\approx \sqrt{u_{fv}}+\dfrac{C_0}{2 \sqrt{u_{fv}}} e^{-C_1 t},\qquad \dfrac{C_0e^{-C_1 \bar{t}}}{u_{fv}}\ll 1.
\eea 
The result is a function $F(t, C_0, C_1, u_{fv})$
\bea
F(t, C_0, C_1, u_{fv})\equiv 1+\dfrac{t^2}{6 \xi u_{fv}}\left(\dot{\phi}-2 V(\phi)\right)
\eea
whose zeros separate the region in which a bounce is allowed from the one in which it is not. As it is to be valid for all values of $t$ such that the asymptotic bounce is reached, we evaluate $F$ at the transition time $\bar{t}$. As $\bar{t}$ depends on the dynamics, we estimate it by its lower bound $C_1^{-1}$, which roughly marks the time at which the mass term dominates the potential. At such time, $\ell_m=C_0 e^{-1}$. As $F$ is a decreasing function of time, 
\bea
F(t, C_0, C_1, u_{fv})\leq F(C_1^{-1}, C_0, C_1, u_{fv})\qquad \text{for}\qquad t\geq C_1^{-1}.
\eea
Thus, the range of $C_1, C_0$ and $u_{fv}$ for which $F>0$ when evaluated at $\bar{t}=C_1^{-1}$ consist in a wider region than the one for generic $F(t, C_0, C_1, u_{fv})$. One gets
\bea
F(x,\xi)\equiv F(C_1^{-1}, C_0, C_1, u_{fv})=\dfrac{-8 (6 \xi +1)+8 (6 \xi +1) (x+1)^{3/2}-3 x (8 \xi  (x+3)+x+4)}{24 \xi  (x+1)}+1\notag\\
 \eea
where $ x=\dfrac{C_0}{u_{fv} e}$. The zeros of this function are reported in Fig.\ref{fig:gplot}. For $x$s below the curve the bounce is allowed, for values above it it is forbidden. As mentioned above, the allowed values of $x$ for arbitrary $\bar{t}$ are actually even smaller then the ones reported in Fig.\ref{fig:gplot}, thus implying even larger $\phi_{\rm fv}$ in order to have a bounce for a given $\ell_m$.\\

In conclusion, there is an upper bound on $\dfrac{\ell_m}{\phi_{\rm fv}}$ (setting $\ell_m=C_0 e^{-1}$) for which a bounce exists. In spite being formally similar to the result found in the massless case (see Table \ref{tab:prevres}), it is physically is very different. First, of all, such bound was derived for values of $t$ such that the scalar field has not yet reached the asymptotic bounce, in contrast to what is described here. Moreover, if the scalar field mass is much smaller than the barrier width, the bound on $\phi_{\rm fv}$ given by the former is accordingly milder than the one given by the latter. This implies that the mass term in the Higgs potential is effectively negligible for our considerations as its mass is much smaller than the barrier width.\\

\begin{figure}
   \mbox{
   \begin{minipage}{0.5\textwidth}
   \includegraphics[scale=0.62]{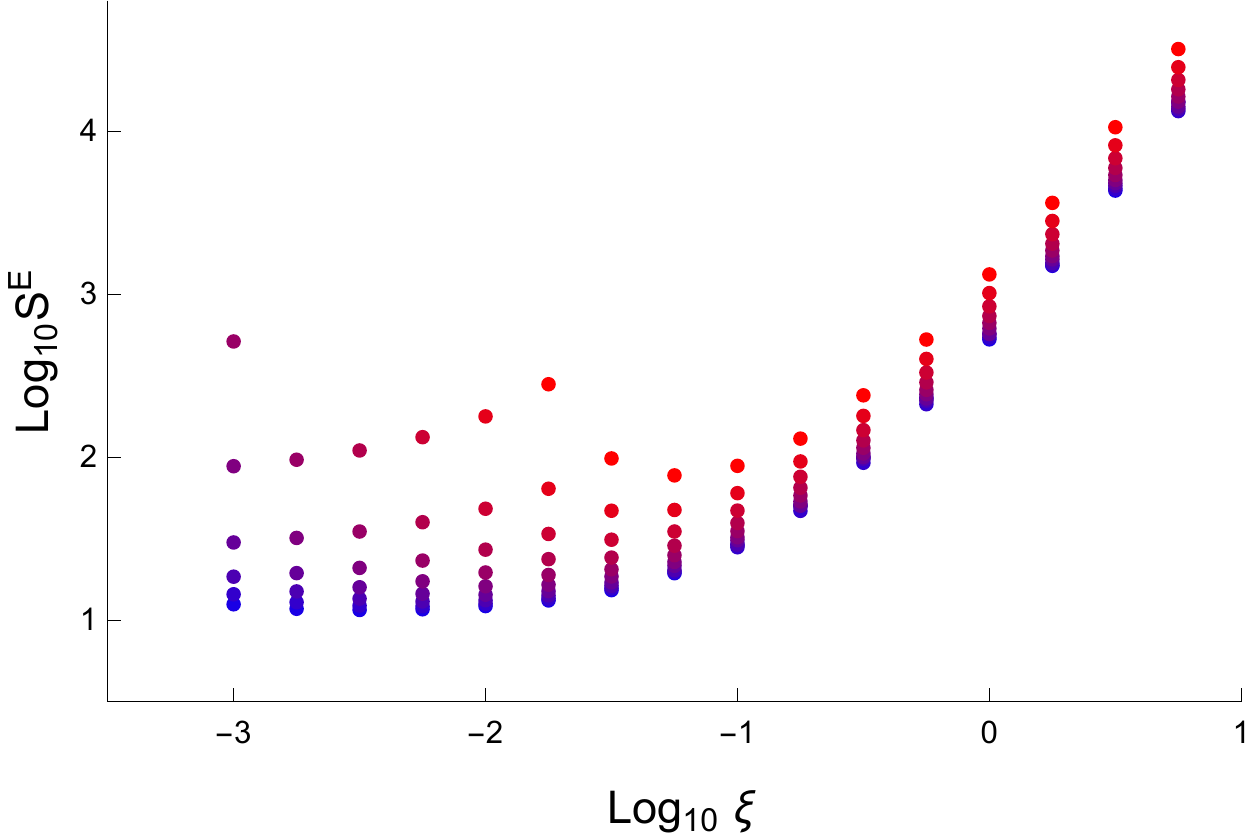}
   \end{minipage}
   \begin{minipage}{0.5\textwidth}
    \includegraphics[scale=0.65]{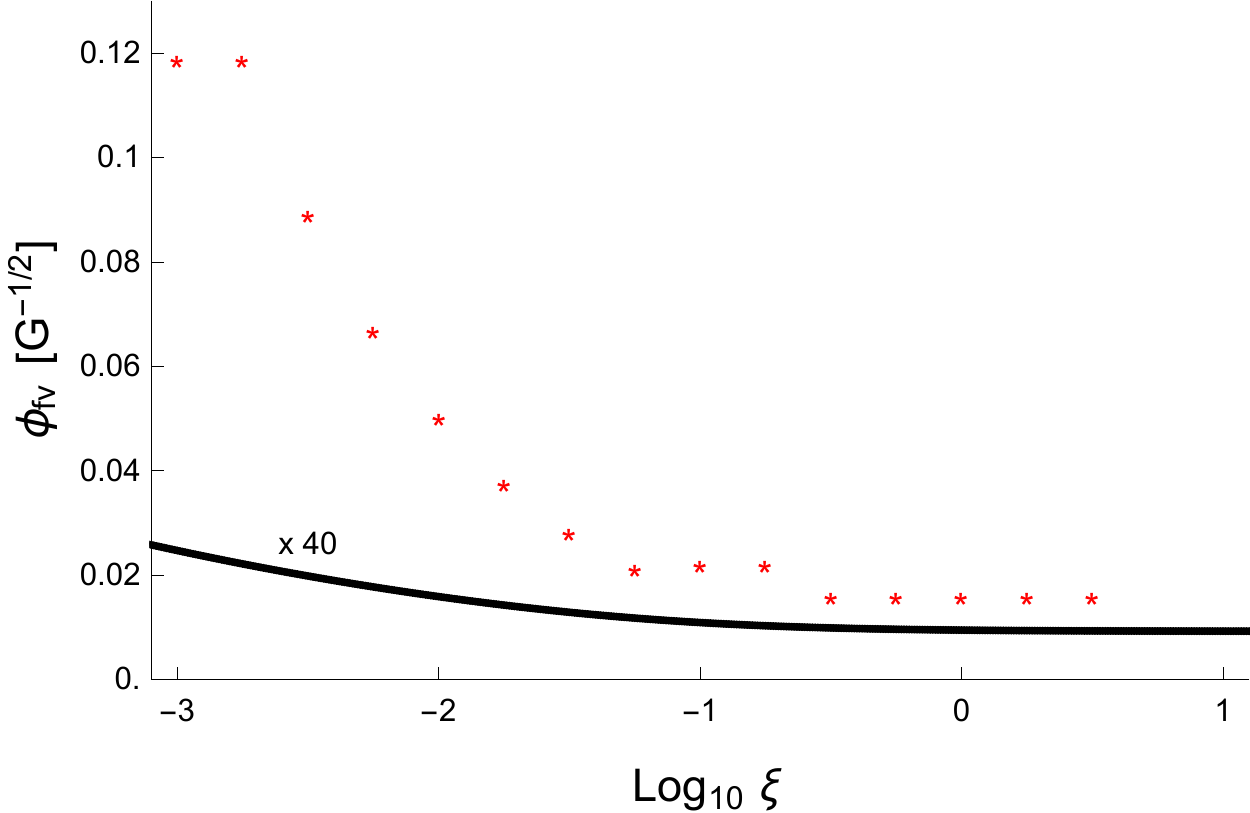}
   \end{minipage}
  }
    \caption{Left: numerical bounce action as a function of  $x$ (in units $G=1$). $\phi_{\rm fv}$ is changed from $\phi_{\rm fv}=0.5$ (blue) to $\phi_{\rm fv}=0.016$ (red). The bounce disappears for increasingly lower values of $\phi_{\rm fv}$ at higher $\xi$. Right: zeros of $F(\phi_{\rm fv},\xi)$ (magnified by a factor 40) as a function of $\xi$ for the potential Eq.\eqref{eq:masspot} (black line) compared with the numerical values (red stars).}
    \label{fig:fvcond}
    \end{figure}

We tested our findings on a scalar field theory with non-minimal coupling to gravity and potential 
  \bea
  \label{eq:masspot}
  V(\phi)=\dfrac{m^2}{2}(\phi-\phi_{\rm fv})^2+g (\phi-\phi_{\rm fv})^3+\lambda (\phi-\phi_{\rm fv})^4
  \eea 
  where $ m^2=3\times 10^{-4},\,g=0.07,\,\lambda=2.25$.
   We found the bounce numerically, varying $\phi_{\rm fv}$ in the interval $[10^{-2}, 0.5]$ and $\xi$ in $[10^{-3}, 10]$ and computing the bounce action as a function of $\xi$. As can be seen in  Fig.\ref{fig:fvcond}, on the left, the bounce disappears at increasingly lower $\phi_{\rm fv}$ for growing values of $\xi$, in agreement with our prediction. We also plotted (see Fig.\ref{fig:fvcond}, on the right) such limiting values of $\phi_{\rm fv}$ as a function of $\xi$ and compared them to the theoretical prediction given by the zeros of $ F(C_1^{-1}, C_0, C_1, u_{fv})$. We estimated $C_0$ by the scalar field value at $\bar{t}^{-1}=C_1$ such that the dominant (90\%) contribution to the potential is given by the mass term, namely we set
  \bea
  \dfrac{2V(\phi)}{m^2 (\phi-\phi_{\rm fv})^2}-1=0.1\qquad \text{with $\phi=\phi_{\rm fv}+C_0 e^{-1}$}
  \eea
  which gives $C_0=6\times 10^{-4}$.
 The numerical values lie well above the theoretical prediction, thus suggesting that $\bar{t}\gg C_1^{-1}$.\\
  \subsection{Quadratic gravity}
  We consider here quadratic corrections in the scale invariant limit, namely we set $\alpha\neq 0$, $M_P^2=0$ and $\xi=0$. As in Sec.\ref{sec:scaledep}, the scalar field on the bounce at large times is independent on $R$, and so it decreases exponentially according to Eq.\eqref{eq:bouncemass}. The solution to the trace equation when the scalar field is given by the asymptotic bounce and Eq.\eqref{eq:approxrho} holds is
      \bea
      \label{eq:Rsiquadratic}
      R(t)=C_1+\dfrac{C_2}{2 t^2}+O(e^{-2 m t})
      \eea
      with $C_1, C_2$ some real constants. Using  Eq.\eqref{eq:Rsiquadratic}, Eq.\eqref{eq:behaviourtp} and taking $\rho(t)=t$ in Eq.\eqref{eq:eom1mod} we find $\dot{\rho}\neq 1$ at large times on the bounce and thus Condition 1. is violated. Thus, there is no bounce for scale invariant gravity with a quadratic term and flat Euclidean spacetime in the false vacuum, as in the massless case.

   \subsection{Non-minimal coupling and quadratic gravity}
For the case $\xi\neq 0$ and $\alpha\neq 0$ we repeat the calculations of Sec.\ref{sec:scaledep} taking $M_P^2=0$. Analogous calculations to the ones reported in Sect.\ref{sec:nonmflat} allows determining the asymptotic bounce.
In order to have a finite bounce action, $f(t)$ should satisfy
\bea
\lim_{t\rightarrow+\infty}  \phi^2 f(t)=0
\eea
as $F(t)$ is monotonically decreasing at large times on the bounce
\bea
f(t)&=&\dfrac{\int_t^{+\infty} t'^{-3} \int_{t'}^{+\infty} t''^{3} F(t'')}{F(t)}\\\non
&\leq& \int_t^{+\infty} t'^{-3} \int_{t'}^{+\infty} t''^{3} =\dfrac{t^2}{8}.
\eea
Then
\bea
 F_1(\phi,t)\approx F_2(\phi,t)\approx F_3(\phi,t)\approx 1
 \eea
 so again 
\bea
 \ddot{\phi}+\dfrac{3 \dot{\phi}}{t}=\dfrac{dV}{d\phi}
 \eea
 for small $\phi$. Then $\phi$ satisfies Eq.\eqref{eq:bouncemass}, at large times on the bounce. This gives (from Eq.\eqref{eq:solnonm})  
 \bea
\label{eq:solscinv}
R=\dfrac{C_1}{2 t^2}+\text{higher orders}
\eea
where higher orders are as in the previous section. Plugging these solutions in Eq.\eqref{eq:eom1mod} one finds that Condition 1. is violated. Setting $\phi_{\rm fv}\neq 0$ amounts to adding a linear non-minimal coupling to gravity, $\phi R$, and having an Einstein-Hilbert-like term on the bounce at large times. Now, though, the "Planck mass" $\sqrt{\xi} \phi_{\rm fv}$ might be of the same order of $m$ and, thus, the considerations that allowed to exclude a bounce in Sec.\ref{sec:nonmflat} do not apply. Thus, we cannot exclude a bounce in this scenario. 

 \section{Conclusion}
 \label{sec:concl}
False vacuum decay in field theory may be formulated as a boundary value problem in Euclidean space. Its solution is called bounce and it is an interpolating trajectory between the tunneling point and the false vacuum. The behaviour near the false vacuum (which is reached at large values of the Euclidean radius) may be found analytically and allows to test for two existence Conditions (Conditions 1. and 2. in the Introduction) for vacuum decay. In this paper, we found the asymptotic bounce in a variety of scalar field theories interacting with gravity. We first focused on single scalar field theories with Einstein-Hilbert gravity and found that massive scalar fields decrease exponentially towards the false vacuum, consistently with previous results in the literature \cite{Affleck:1980mp}. We also found that higher-order kinetic terms do not influence the asymptotic bounce, which is Eq.\eqref{eq:behaviourtp} if the scalar field is massless with small cubic self-interactions and Eq.\eqref{eq:bouncemass} if the scalar field is massive. We determined the asymptotic bounce in a number of spacetime dimensions other than four, finding similar results as in the four dimensional case. The aforementioned theories did not show any violation according to Conditions 1. and 2. Then we focused on massive scalar fields with modified gravity consisting in an Einstein-Hilbert term, a non-minimal coupling and a quadratic Ricci scalar. We found that:
\begin{itemize}
    \item if $M_P\neq 0$, a bounce is allowed when a non-minimal coupling is included, while forbidden with a squared Ricci term;
    \item if $M_P= 0$ a bounce is allowed when a non-minimal coupling is included if the field acquires a non-vanishing vacuum value $\phi_{\rm fv}$ roughly larger than the scalar field value for which the mass term dominates the potential. A bounce is also allowed if we consider both quadratic gravity and a non-minimal coupling, if the scalar field acquires a non-vanishing vacuum value and in a restricted class of theories. In this case the on-shell action may be small if $\phi_{\rm fv}$ is small. Turning off the non-minimal coupling, instead, makes the bounce disappear.
\end{itemize}
The same analysis carried out for massless fields in \cite{nostro} indicates that a mass term slightly widens the set of theories for which a bounce is allowed. 
Then, quadratic Ricci terms forbid a bounce for both massless and massive scalar fields in most scenarios, and, then, also for the Higgs field, independently on the masslessness approximation usually adopted in false vacuum decay calculations. In the future we would like to extend our analysis to de Sitter vacua, in order to explore metastability of the Higgs field in the early Universe
\cite{Rinaldi:2015uvu,Tambalo:2016eqr,Vicentini:2019etr}.

\subsection*{Acknowledgments}

 S.\ V.\ acknowledges the financial support of the Italian National Institute for Nuclear Physics (INFN) for her Doctoral studies. This work has been partially performed using the software Mathematica.
  
\subsection*{Data availability statement}

 All data generated or analysed during this study are included in this published article.
  \appendix
 \section{Time-dependent Taylor expansion of a radius-dependent potential}
 \label{sec:appendix}
Here we compute the coefficients $f_n$ of Eq.\eqref{eq:taylorV} in the case of a radius-dependent potential. We denote (total) derivatives of arbitrary order  with the index $(n)$, while derivatives of first and second order of the scalar field are denoted by one dot or two dots respectively. Partial derivatives in the scalar field and in $t$ are denoted with the symbol $\partial$. Partial derivatives of order $i$ are indicated as  $\partial^i$. The equations are implicitly evaluated at $t^*$ (such that $\dot{\phi}(t^*)=0$)  and field value $\phi_*$. 
Using the equation of motion for the scalar field we can write
 \begin{equation}
 \label{eq:app_exp_eom1}
    \dfrac{\partial^iV}{\partial\phi^i}^{(n+1)}=\left(  \dfrac{\partial^{i+1}V}{\partial\phi^{i+1}}\dot{\phi}+\dfrac{\partial^{i+1}V}{\partial t\partial\phi^{i}}\right)^{(n)}\,,\qquad \phi^{(n)}=\left(  \dfrac{\partial^2V}{\partial\phi^2}\dot{\phi}+\dfrac{\partial^2V}{\partial\phi \partial t}\right)^{(n-3)}+\sum_{i=2}^{n-1}B_i \dfrac{\phi^{(i)}}{(t+a)^{n-i}}
 \end{equation}
 where $B_i$s are numerical factors, whose value is not relevant for the following discussion. As the explicit $t$ dependence of the potential makes the calculation more involved with respect to the radius independent case (reported in \cite{nostro}), we omit numerical coefficients for simplicity in the following.
 Using the first equation in  
 \eqref{eq:app_exp_eom1}, we can write the (n+1)-th derivative of $\dfrac{dV}{d\phi}$ as 
 \begin{equation}
 \begin{split}
   \dfrac{\partial V}{\partial\phi}^{(n+1)}=& \dfrac{\partial^2V}{\partial\phi^2}\phi^{(n+1)}+\dots+   \left(\dfrac{\partial^2V}{\partial\phi^2}\right)^{(n-1)}\ddot{\phi}+\dfrac{\partial^2V}{\partial\phi \partial t}^{(n)}=\ddot{\phi}\left( \left(\dfrac{\partial^3V}{\partial\phi^3}\right)^{(n-3)}\ddot{\phi}+\dots+\right.\\&+\left. \dfrac{\partial^3V}{\partial\phi^3}\phi^{(n-1)}\right)+\phi^{(3)}\left( \left(\dfrac{\partial^3V}{\partial\phi^3}\right)^{(n-4)}\ddot{\phi}+\dots+ \dfrac{\partial^3V}{\partial\phi^3}\phi^{(n-2)}\right)+\dots+\dfrac{\partial^2V}{\partial\phi^2}\phi^{(n+1)}+\\&+\dfrac{\partial}{\partial t}\left(\dfrac{\partial^2V}{\partial\phi^2}\phi^{(n)}+\dots+\dfrac{\partial^2V}{\partial\phi^2}^{(n-2)}\ddot{\phi}+\dfrac{\partial^2V}{\partial\phi \partial t}^{(n-1)}\right)+\sum_{j=1}^{n-1}\dfrac{\partial^2V}{ \partial t\partial\phi^2}^{(n-j-1)}\phi^{(j+1)},
  \end{split}
  \end{equation}
  which can be further expanded using again Eq.\eqref{eq:app_exp_eom1}. We obtain
 \begin{equation}
 \begin{split}
 \label{eq:appendix_exp}
     \dfrac{\partial V}{\partial\phi}^{(n+1)}=&\dfrac{\partial V}{\partial \phi \partial t^{n+1}}+ \sum_j\dfrac{\partial^2V}{\partial\phi^2\partial t^j}\phi^{(n+1-j)}+\sum_j\dfrac{\partial^3V}{\partial\phi^3\partial t^j}(\ddot{\phi}\,\phi^{(n-1-j)}+\phi^{(3)} \phi^{(n-2-j)}+\dots+\\&+\phi^{(n+1)/2}\phi^{(n+1)/2-j})+\sum_j\dfrac{\partial^4V}{\partial\phi^4\partial t^j}(\ddot{\phi}^2\phi^{(n-3-j)}+\phi^{(3)}\ddot{\phi} \phi^{(n-4-j)}+\dots+\\&+\phi^{(n+1)/3}\phi^{(n+1)/3}\phi^{(n+1)/3-j})+\dots.
  \end{split}\end{equation}
 where sums on $j$ run from $j=0$ to some upper limit, for which derivatives of the scalar field are of order two.
The result is similar to the $t-$independent case: each term $\dfrac{\partial^{i+j}
V}{\partial\phi^i\partial t^j}$ in Eq.\eqref{eq:appendix_exp} is multiplied by $i-1$ terms, which are derivatives of $\ddot{\phi}$. Such derivatives are of order $n+5-2i$ or lower, thus these terms are non-vanishing only if $n+5-2i>1$. So, the highest-order derivative $\dfrac{\partial^{\bar{\imath}}V}{\partial\phi^{\bar{\imath}}}$ that appears in  Eq.\eqref{eq:appendix_exp} is the one satisfying $n+5-2\,\bar{\imath}=3$ for even $n$ and $n+5-2\,\bar{\imath}=2$ for odd $n$.  The difference with respect to the $t-$independent case is that now the potential contains also partial derivatives in $t$ and such terms are multiplied by $i-1$ terms, which are derivatives of $\ddot{\phi}$, the only exception being $t$ derivatives of $\dfrac{\partial V}{\partial \phi}$. For example, the sixth-order derivative $\dfrac{\partial V}{\partial\phi}^{(6)}$ ($n=5$) is expanded in terms of $\dfrac{\partial ^2V}{\partial\phi^2}$, $\dfrac{\partial^3V}{\partial\phi^3}$ and $\dfrac{\partial^4V}{\partial\phi^4}$ as: 
 \begin{equation}
    \begin{split}
         \dfrac{\partial V}{\partial\phi}^{(6)}=&\dfrac{\partial^2V}{\partial\phi^2}\phi^{(6)}+\dfrac{\partial^2V}{\partial\phi^2\partial t}\phi^{(5)}+\dfrac{\partial^2V}{\partial\phi^2\partial t^2}\phi^{(4)}+\dfrac{\partial^2V}{\partial\phi^2\partial t^3}\phi^{(3)}+\dfrac{\partial^2V}{\partial\phi^2\partial t^4}\phi^{(2)}+\dfrac{\partial^3V}{\partial\phi^3}(\ddot{\phi}\,\phi^{(4)}+\\&+ \phi^{(3)} \phi^{(3)})+\dfrac{\partial^3V}{\partial\phi^3\partial t}\ddot{\phi}\,\phi^{(3)}+\dfrac{\partial^3V}{\partial\phi^3\partial t^2}\ddot{\phi}^2+\dfrac{\partial^4V}{\partial\phi^4} \ddot{\phi}^3+\dfrac{\partial V}{\partial\phi \partial t^6}.
     \end{split}
 \end{equation}
 
 We expand  derivatives of $\ddot{\phi}$ in Eq.s\eqref{eq:app_exp_eom1} using Eq.\eqref{eq:appendix_exp}. We find 
 \begin{equation}
     \begin{split}
 \label{app_der_phi}
     \phi^{(n+1)}=&\sum_j\dfrac{\partial^2V}{\partial\phi^2\partial
     t^j}\phi^{(n-1-j)}+\sum_j\dfrac{\partial^3V}{\partial\phi^3\partial t^j}(\phi^{(n-3-j)}\ddot{\phi}+\phi^{(n-4-j)}\phi^{(3)}+\dots)+\\&+\sum_{i=2}^{n}B_i \dfrac{\phi^{(i)}}{(t+a)^{n-i+1}}+\dfrac{\partial V}{\partial\phi \partial t^{n-1}}
     \end{split}
 \end{equation}
 As a result, using Eq.\eqref{app_der_phi}, we can express Eq.\eqref{eq:appendix_exp} in terms of partial derivatives of the potential with respect to the scalar field and $t$, $\ddot{\phi}$ and $t^*$ only.  We order such terms according to the order of the derivative of the potential in the scalar field, which will be labelled with $i$ in the following. Partial derivatives in $t$ of the potential should be compensated with appropriate powers of $t^*$ with respect to the $j=0$ term. Moreover, $t$ derivatives of $\dfrac{\partial  V}{\partial \phi }$ compensate for some factors  $\dfrac{\ddot{\phi}}{t^j}$ with respect to the $t-$independent case, as can be seen from Eq.\eqref{app_der_phi}. We can carry out the calculation in the $t-$independent case, as explained in \cite{nostro} , and then add terms $\dfrac{\partial^{i+j}
V}{\partial\phi^i\partial t^j} t^j$ to $\dfrac{\partial^{i}
V}{\partial\phi^i}$ and $\dfrac{\partial  V}{\partial \phi \partial  t^j }$ to $\dfrac{\ddot{\phi}}{t^j}$. All such terms that are multiplied by a negative or vanishing power of $t^*$ contribute. It is easier to see how this works with some examples. The highest-order derivative (the $\bar{\imath}$-th term) is multiplied only by $t$ derivatives of the scalar field of order $2$ or $3$ and thus it contributes as
 \begin{gather} 
    \dfrac{\partial^{\bar{i}}V}{\partial\phi^{\bar{\imath}}}\ddot{\phi}^{\bar{\imath}-1}\quad\text{odd $n$,}\\
    \qquad\dfrac{\ddot{\phi}^{\bar{\imath}-1}}{(t^*+a)}\left(\dfrac{\partial^{\bar{\imath}}V}{\partial\phi^{\bar{\imath}}}+
   \dfrac{\partial^{\bar{\imath}}V}{\partial\phi^{\bar{\imath}}\partial t}(t^*+a)\right)+\ddot{\phi}^{\bar{\imath}-2}\dfrac{\partial V}{\partial\phi \partial t} \dfrac{\partial^{\bar{\imath}}V}{\partial\phi^{\bar{\imath}}}\quad\text{even $n$}
 \end{gather}
 to $f_n$. The first term for even and odd $n$ is present also in the $t-$independent case: two additional terms appear, one that replaces $\dfrac{\ddot{\phi}}{t^*}$ and a $t$ derivative which is compensated by an additional power of time. As positive powers of $t$ cannot appear, there should be no other terms in the highest-order derivative.\\
 The second-highest derivative $\bar{\imath}-1$ is multiplied by derivatives of the scalar field of order $2,\,3,\,4,\,5$. Using Eq.\eqref{app_der_phi}, derivatives of order $4$ and $5$ may be expressed in terms of lower order derivatives. We get
 \begin{equation}
      \begin{split}\text{odd $n$}\qquad&\dfrac{\ddot{\phi}^{\bar{\imath}-2}}{(t^*+a)^2}\left[\dfrac{\partial^{\bar{\imath}-1}V}{\partial\phi^{\bar{\imath}-1}}\left(1+\dfrac{\partial^2V}{\partial\phi^2} (t^*+a)^2\right)+\dfrac{\partial^{\bar{\imath}}V}{ \partial\phi^{\bar{\imath}-1}\partial t}(t^*+a)+\dfrac{\partial^{\bar{\imath}+1}V}{\partial\phi^{\bar{\imath}-1}\partial t^2}(t^*+a)^2\right]+\\&+\dfrac{\ddot{\phi}^{\bar{\imath}-3}}{(t^*+a)}\left(\dfrac{\partial^{\bar{\imath}-1}V}{\partial\phi^{\bar{\imath}-1}}\dfrac{\partial V}{\partial\phi \partial t}+\dfrac{\partial^{\bar{\imath}-1}V}{\partial\phi^{\bar{\imath}-1}}\dfrac{\partial V}{\partial\phi  \partial t^2}(t^*+a)+\dfrac{\partial^{\bar{\imath}}V}{\partial\phi^{\bar{\imath}-1}\partial t}\dfrac{\partial V}{\partial\phi \partial t}(t^*+a)\right)+\\&+\dfrac{\partial^{\bar{\imath}-1}V}{\partial\phi^{\bar{\imath}-1}}\ddot{\phi}^{\bar{\imath}-4}\dfrac{\partial V}{\partial\phi \partial t}\dfrac{\partial V}{\partial \phi \partial t}
     \end{split}
 \end{equation} 
 \begin{equation}
     \begin{split}\text{even $n$}\qquad&\dfrac{\ddot{\phi}^{\bar{\imath}-2}}{(t^*+a)^3}\left[\dfrac{\partial^{\bar{\imath}-1}V}{\partial\phi^{\bar{\imath}-1}}\left(1+\dfrac{\partial^2V}{\partial\phi^2} (t^*+a)^2+\dfrac{\partial^3V}{\partial\phi^2\partial t } (t^*+a)^3\right)+\dfrac{\partial^{\bar{\imath}}V}{\partial\phi^{\bar{\imath}}\partial t}(t^*+a)\right.\\&\left.\left(1+\dfrac{\partial^2V}{\partial\phi^2} (t^*+a)^2\right)+\dfrac{\partial^{\bar{\imath}+1}V}{\partial\phi^{\bar{\imath}-1}\partial t^2}(t^*+a)^2+\dfrac{\partial^{\bar{\imath}+2}V}{\partial\phi^{\bar{\imath}-1}\partial t^3}(t^*+a)^3\right]+\\&+\dfrac{\ddot{\phi}^{\bar{\imath}-3}}{(t^*+a)^2}\left[\dfrac{\partial^{\bar{\imath}-1}V}{\partial\phi^{\bar{\imath}-1}}\dfrac{\partial V}{\partial\phi \partial t}\left(1+\dfrac{\partial^2V}{\partial\phi^2} (t^*+a)^2\right)+\dfrac{\partial^{\bar{\imath}-1}V}{\partial\phi^{\bar{\imath}-1}}\dfrac{\partial V}{\partial\phi \partial t^2}(t^*+a)+\right.\\&\left.+\dfrac{\partial^{\bar{\imath}}V}{\partial\phi^{\bar{\imath}-1}\partial t}\dfrac{\partial V}{\partial\phi \partial t}(t^*+a)+\dfrac{\partial^{\bar{\imath}+1}V}{\partial\phi^{\bar{\imath}-1}\partial t^2}\dfrac{\partial V}{\partial\phi dt}+\dfrac{\partial^{\bar{\imath}}V}{\partial\phi^{\bar{\imath}-1}\partial t}\dfrac{\partial V}{\partial\phi\partial t^2}+\dfrac{\partial^{\bar{\imath}-1}V}{\partial\phi^{\bar{\imath}-1}}\dfrac{\partial V}{\partial\phi \partial t^3}\right]+\\&+\dfrac{\ddot{\phi}^{\bar{\imath}-4}}{(t^*+a)}\left[\dfrac{\partial^{\bar{\imath}-1}V}{\partial \phi^{\bar{\imath}-1}}\dfrac{\partial V}{\partial\phi \partial t}\dfrac{\partial V}{\partial \phi \partial t}+\dfrac{\partial ^{\bar{\imath}}V}{\partial\phi^{\bar{\imath}-1}\partial t}\dfrac{\partial V}{\partial \phi \partial t}\dfrac{\partial V}{\partial \phi \partial t}(t^*+a)+\right.\\&\left.\dfrac{\partial^{\bar{\imath}-1}V}{\partial\phi^{\bar{\imath}-1}\partial t}\dfrac{\partial V}{\partial \phi \partial t^2}\dfrac{\partial V}{\partial\phi \partial t}(t^*+a)\right]\end{split}\end{equation}
   The first term for even and odd $n$ are the same as in the $t-$independent case, and other contributions appear with additional derivatives, following the rules described above. Terms that appeared as 
  \bea
  \dfrac{\partial^iV}{\partial\phi^i} t^{2i-2} \ddot{\phi}^{i-2} 
  \eea
  in the radius independent case are now
  \bea
  \dfrac{\partial^{i}V}{\partial\phi^i\partial t^j} t^{2i-2+j}\ddot{\phi}^{i-2}.
  \eea
If they are finite in the large $t^*$ limit one gets, apart from numerical factors,
\begin{equation}
\begin{split}
 &\dfrac{\partial V}{\partial\phi}^{(n+1)}=\\&=\dfrac{\partial ^{n+1}V}{\partial \phi \partial t^{n+1}}+\sum_{i=0}^{i=\bar{\imath}-2}\,\sum_{m=0}^{\bar{\imath}-i-1}\sum_{j_0,\dots j_m}\sum_{n=0}^{2i+1}\dfrac{\partial V^{\bar{\imath}-i}}{\partial \phi^{\bar{\imath}-i}\partial t^{j_0}} \dfrac{\ddot{\phi}^{\bar{\imath}-i-1-m}}{t^{2i+1-n}}\dfrac{\partial V}{\partial \phi \partial t^{j_1}}\times\dots \times \dfrac{\partial V}{\partial \phi \partial t^{j_m}}
 \end{split}
 \end{equation}
where $j_0+\dots+j_m=n$, while in the radius independent case one gets
\bea
 \dfrac{\partial V}{\partial \phi}^{(n+1)}=\sum_{i=0}^{i=\bar{\imath}-2}\dfrac{\partial V^{\bar{\imath}-i}}{\partial\phi^{\bar{\imath}-i}} \dfrac{\ddot{\phi}^{\bar{\imath}-i-1}}{t^{2i+1}}.
\eea 
 Imposing also that
    \bea
  \dfrac{\partial^iV}{\partial\phi^i\partial t^j}t^j\ll\dfrac{\partial^iV}{\partial\phi^i},
  \eea
  and that
  \bea
  \label{eq:timecond2}
 \dfrac{\partial^jV}{\partial\phi \partial t^j}t^j\ll\ddot{\phi},
  \eea
  besides Eq.\eqref{eq:massless}, one has that Eq.\eqref{eq:behaviourtp} is satisfied in the radius-dependent case. In the massive case the only non vanishing derivative of the potential in $\phi$ is of order two and
  \bea
  \dfrac{\partial^2V}{\partial\phi^2} t^2\rightarrow +\infty
  \eea
  while 
   \bea
  \dfrac{\partial^2V}{\partial\phi^2\partial t^j} t^j \approx t^{-2}
  \eea
  and
  \bea
   \dfrac{\partial^jV}{\partial\phi \partial t^j}t^j\approx \phi\,t^{-2}
  \eea
  then, the mass term dominate over the radius dependent ones, and they are important at each order in the Taylor expansion thus giving Eq.\eqref{eq:eommass}.

\end{document}